\documentclass[journal=aaembp,manuscript=article]{achemso}

\usepackage[version=3]{mhchem} 

\author{Alexandros Bampis}
\affiliation{Institute of Physics, Ecole Polytechnique Fédérale de Lausanne (EPFL), CH-1015 Lausanne, Switzerland}
\email{alexandros.bampis@epfl.ch}
\author{Johann Stachurski}
\affiliation{Institute of Physics, Ecole Polytechnique Fédérale de Lausanne (EPFL), CH-1015 Lausanne, Switzerland}
\author{Anna Schwab}
\affiliation{Institute of Physics, Ecole Polytechnique Fédérale de Lausanne (EPFL), CH-1015 Lausanne, Switzerland}
\author{Jean-François Carlin}
\affiliation{Institute of Physics, Ecole Polytechnique Fédérale de Lausanne (EPFL), CH-1015 Lausanne, Switzerland}
\author{Raphaël Butté}
\affiliation{Institute of Physics, Ecole Polytechnique Fédérale de Lausanne (EPFL), CH-1015 Lausanne, Switzerland}
\author{Nicolas Grandjean}
\affiliation{Institute of Physics, Ecole Polytechnique Fédérale de Lausanne (EPFL), CH-1015 Lausanne, Switzerland}

\title{Depth Control of Room-Temperature Quantum Emitters in Gallium Nitride}


\begin{document}

\begin{abstract}
    Bright quantum emitters are key components for quantum communication systems. Radiative point defects in gallium nitride (GaN) are promising candidates for room-temperature single-photon emission, operating from the visible to the telecom O-band. Despite their potential, their integration into photonic structures has remained limited in the literature, with experimental photon extraction efficiencies far below simulated predictions. To identify the origin of this limitation, we investigate visible and near-infrared quantum emitters in GaN epilayers grown on $c$-plane sapphire substrates exhibiting narrow linewidths ($\sim$4 nm), high photon count rates ($>$2 MHz), and strong antibunching, reaching $g^{(2)}(0)$ values as low as 0.06 at room temperature. We find that these emitters are located near the GaN/substrate interface, explaining their limited coupling to optical modes. Building on this observation, we show that the insertion of a thin low-temperature GaN interlayer enables the formation of quantum emitters at arbitrary depths with sub-60 nm accuracy, independent of the substrate. The intentionally introduced emitters retain optical properties comparable to naturally occurring ones, including narrow linewidths ($\sim$6 nm), saturation count rates exceeding 1.5 MHz, high Debye-Waller factors (0.69-0.98), and strong antibunching. The resulting epilayer fully coalesces within less than 250 nm ensuring compatibility with GaN-based cavity fabrication and enabling emitter placement within intrinsic regions of p-i-n diode architectures. These results mark a decisive step toward efficient emitter-cavity coupling, enabling the realization of cavity-enhanced quantum emission in GaN.
\end{abstract}

\section{Introduction}

The realization of robust quantum communication networks is critically dependent on the development of efficient and scalable single-photon sources \cite{Kimble2008}. Semiconductor quantum dots based on III-arsenides \cite{Quantum_boxes_marzin_1994, Rastelli_gaas_qds} have been extensively developed and serve as nearly ideal single-photon sources \cite{Senellart2017, schweickert_2017}. However, their optimal performance requires cryogenic temperatures, posing a significant challenge to their large-scale implementation in practical quantum networks. In contrast, III-nitride-based quantum dots can operate up to room temperature \cite{holmes_gan_qds, Sebastian_GaN_QD}, but their short emission wavelengths ($\lambda<360$ nm) are not suitable for long distance quantum communication. As a promising alternative, point defects in wide-bandgap semiconductors such as diamond \cite{Kurtsiefer2000}, silicon carbide \cite{Castelletto2014}, hexagonal boron nitride \cite{grosso_tunable_2017, chatterjee_room-temperature_2025} and more recently, gallium nitride (GaN) \cite{Berhane2017, zhou_2018}, have attracted growing attention. These systems offer a unique combination of desirable properties, including sharp zero-phonon lines (ZPLs), large Debye-Waller factors, high brightness and strong antibunching up to room temperature.

GaN, in particular, is an attractive host material owing to mature fabrication processes \cite{houdre_phc_2012}, its bipolar doping capabilities \cite{Nakamura1992} and established industry dedicated to solid-state lighting. Recent studies have identified radiative point defects in GaN that function as bright quantum emitters (QEs) in the visible \cite{Berhane2017}, near-infrared \cite{bishop_SIL_2022} and telecom \cite{zhou_2018} wavelength ranges, exhibiting strong antibunching and high-contrast optically detected magnetic resonance (ODMR) at room temperature \cite{luo_odmr_2024, odmr_telecom_2025}. These figures of merit position GaN-based QEs as promising building blocks for future quantum photonic devices.

However, the nature of these QEs remains elusive. While efforts have been made to characterize their properties \cite{berhane_photophysics_2018, geng_optical_2023, luo_odmr_2024}, much less is known about their formation mechanisms and spatial localization within the GaN crystal. The defects are often considered as forming randomly within the epilayer \cite{nguyen_effects_2019}. This lack of spatial control severely hinders practical deployment, as efficient photon extraction relies on optimal coupling to photonic structures, which in turn requires precise emitter positioning both in-plane and in-depth. A very recent study reported control over the in-depth position of QEs in GaN grown on silicon using specific growth conditions \cite{eggleton_controlled_2026}. This opens new opportunities for the integration of GaN QEs into photonic circuits.

In this work, we focus on visible and near-infrared (NIR) QEs and demonstrate that, when GaN is grown on $c$-plane sapphire, naturally occurring QEs are located in the vicinity of the GaN/sapphire interface. This vertical position inherently limits the optical coupling between the QEs and common photonic structures designed to enhance photon extraction efficiency and/or modify the emitter lifetime via the Purcell effect, such as $\lambda$-cavity micropillars,\cite{somaschi_near-optimal_2016} circular Bragg gratings,\cite{meunier_2023} and waveguides.\cite{lu_bright_2020} We identify the low-temperature nucleation layer as the key growth step associated with QE formation and incorporate an equivalent thin interlayer within GaN buffers, thereby enabling QE formation at selected depths with a vertical accuracy better than 60 nm, independent of the nature of the substrate. The introduced QEs exhibit room-temperature optical properties comparable to naturally occurring emitters, including narrow ZPL linewidths around 6 nm, Debye-Waller factors ranging from 0.69 to 0.98, saturation count rates above 1.5 MHz, and strongly anti-bunched emission. The resulting epilayer enables the controlled placement of emitters within the intrinsic region of p-i-n diode architectures and within GaN-based cavities. Furthermore, the correlation between emitter density and impurity depth profiles obtained by secondary ion mass spectrometry (SIMS) indicates that oxygen may be involved in the nature of these optically active defects, although further investigation is required to confirm this hypothesis.

These results mark a critical step toward understanding the origin of room-temperature QEs in GaN and enable scalable fabrication of quantum photonic devices.

\section{Results and discussion}

\subsection{Optical properties of naturally occurring QEs in GaN on sapphire}

\begin{figure}[h!]
\centering
\includegraphics[width=1.\textwidth]{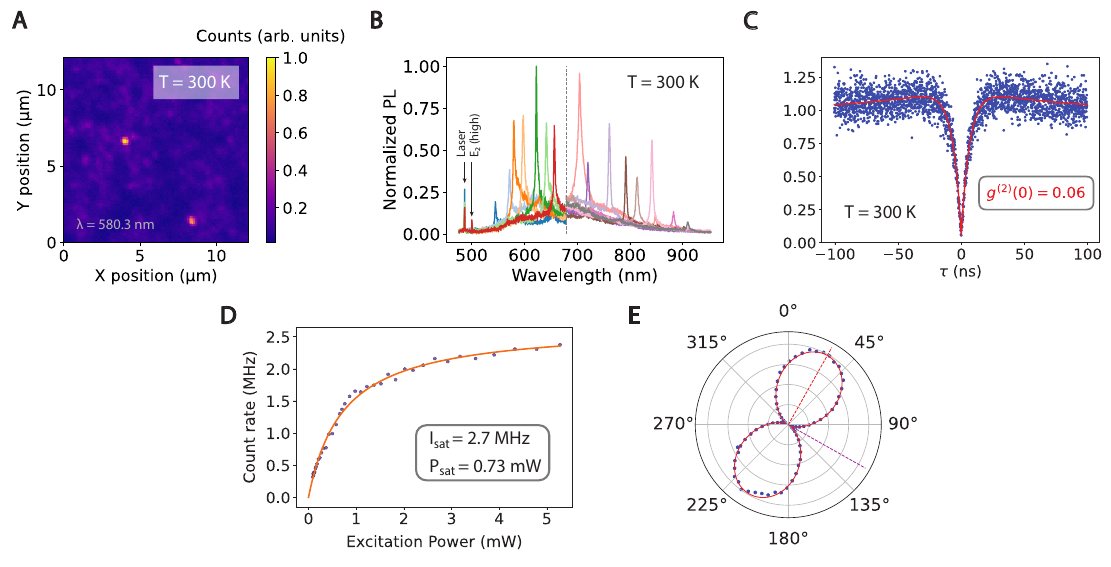}
\caption{Optical properties of naturally occurring emitters in GaN. (A) Room-temperature PL map of a GaN layer grown on $c$-plane sapphire, displaying the normalized PL intensity within a narrow spectral window centered at 580.3 nm. Two QEs emitting at this wavelength appear as bright spots on the map. (B) Optical signatures of 15 distinct QEs with emission wavelengths ranging from 550 nm to 910 nm. The spectra are selected from 2 PL maps acquired with complementary spectral windows (vertical dashed line). (C) Second-order correlation function measurement of a GaN QE at room temperature fitted with a 3-level system model \cite{meunier_2023}, yielding g$^{(2)}(0)=0.06$. The excitation power was set to 12.5 µW in this measurement. (D) Power-dependent photon count rate of a GaN QE, fitted with a 3-level system saturation behavior \cite{Berhane2017}, exhibiting a saturation count rate of I$_{\text{sat}} = 2.7$ MHz. The data are background-substracted with a measurement taken 600 nm away from the QE (laser spot size of 216 nm). (E) Background-substracted, polarization-dependent PL measurement of a GaN QE displaying a 91 \% DLP. The main polarization axis lies at 29° (red dashed line), with the 0° angle indicating the direction perpendicular to the wurtzite $m$-plane. Dipole orientation can vary across emitters\cite{geng_optical_2023}.}\label{Fig_1}
\end{figure}

Figure \ref{Fig_1} summarizes representative room-temperature micro-photoluminescence (µ-PL) measurements performed on GaN-on-sapphire samples grown in-house by metalorganic vapor-phase epitaxy (MOVPE). A continuous-wave (CW) 488 nm laser is used to excite the sample through an oil-immersion objective mounted on an $XYZ$ nanopositioning stage for spatial mapping. The collected photoluminescence signal is directed either to a spectrometer equipped with a charge-coupled device (CCD) for spectral analysis or to a Hanbury Brown and Twiss (HBT) interferometer to measure the photon statistics. In our measurements, µ-PL maps are acquired by recording a full PL spectrum at each spatial position. As a result, the PL maps can be visualized at a selected wavelength or over arbitrary wavelength ranges. In Fig. \ref{Fig_1} (A), for example, the displayed map corresponds to the signal recorded by a single pixel of the spectrometer CCD, corresponding to the signal detected at 580.3 nm with a spectral bandwidth of 0.3 nm. This representation reveals an emitter density that enables the isolation and characterization of individual QEs with appropriate spectral filtering. A panchromatic µ-PL map over the 520-780 nm spectral range is provided in Fig. S2 of the Supporting Information (SI). Over this wavelength range, the QE density is estimated to be on the order of $3 \times 10^{8}$ cm$^{-2}$. Representative emission spectra of 15 QEs are illustrated in Fig. \ref{Fig_1} (B), all exhibiting sharp ZPLs, with a median full width at half maximum (FWHM) of 4.2 nm. The broadband background luminescence visible across all spectra is attributed to the GaN yellow luminescence (YL).\cite{reshchikov_origin_2023} Consistent with earlier reports \cite{Berhane2017, luo_odmr_2024}, these emitters exhibit a broad distribution of emission wavelengths. Notably, our measurements extend the documented emission range from below 550 nm to above 900 nm, limited only by the excitation wavelength and the silicon-based CCD response. This result suggests that GaN QEs could cover a virtually gap-free range of wavelengths all the way to the telecom O-band \cite{zhou_2018, meunier_2023}. 

The quantum nature of these emitters is confirmed through second-order correlation function (\textit{g}$^{(2)}(\tau)$) measurements: for the emitter highlighted in Fig. \ref{Fig_1} (C), we observe excellent antibunching with a \textit{g}$^{(2)}(0)$ value of 0.06 at room temperature. The observed emitters also display high brightness without patterning of any additional photonic structure, with detected saturation count rates exceeding 2.5 MHz (Fig. \ref{Fig_1} (D)) and degrees of linear polarization (DLPs) above 90\% (Fig. \ref{Fig_1} (E)), while exhibiting variability in dipole orientation \cite{geng_optical_2023}. To enable the depth-resolved analyses presented below, the laser excitation line along with the GaN $E_2$(high) Raman mode \cite{Raman_GaN_1995} are intentionally retained in the signal sent to the spectrometer, while these contributions are removed for \textit{g}$^{(2)}(\tau)$, power- and polarization-dependent measurements.

\subsection{Origin of QEs in GaN on sapphire}

Although GaN QEs already exhibit attractive room-temperature properties, their integration into photonic cavities is essential to enhance their emission rate and photon collection efficiency \cite{gerard_1998, purcell_1946, Senellart2017}. Such integration requires precise spectral and spatial matching between the emitter and the cavity mode. Spectral matching can be accomplished through careful design and fabrication of the photonic structure \cite{dousse_controlled_2008, meunier_2023}, post-fabrication tuning of the cavity \cite{krieger_postfabrication_2024, kaupp_postfabrication_2025}, electrical or strain tuning of the QE emission wavelength \cite{wijitpatima_bright_2024, akbari_tunable_hbn_2022, grosso_tunable_2017, yuan_first-principle_2023}, while the in-plane emitter position may be controlled during growth \cite{chen_laser_2017, fournier_position-controlled_2021} or determined prior to cavity fabrication  \cite{sapienza_2015, descamps_2024}. Unlike epitaxial quantum dots, where the emitter depth is precisely controlled and defined during growth, defect-based emitters in GaN on sapphire are expected to exhibit a stochastic depth distribution within the thin film \cite{nguyen_effects_2019}, hindering their efficient coupling to photonic structures. To verify this assumption, we examine the vertical position of naturally occurring QEs using depth-resolved µ-PL. We then demonstrate control over their in-depth position with high precision thanks to epitaxial growth engineering.

\begin{figure}[ht]
\centering
\includegraphics[width=1.\textwidth]{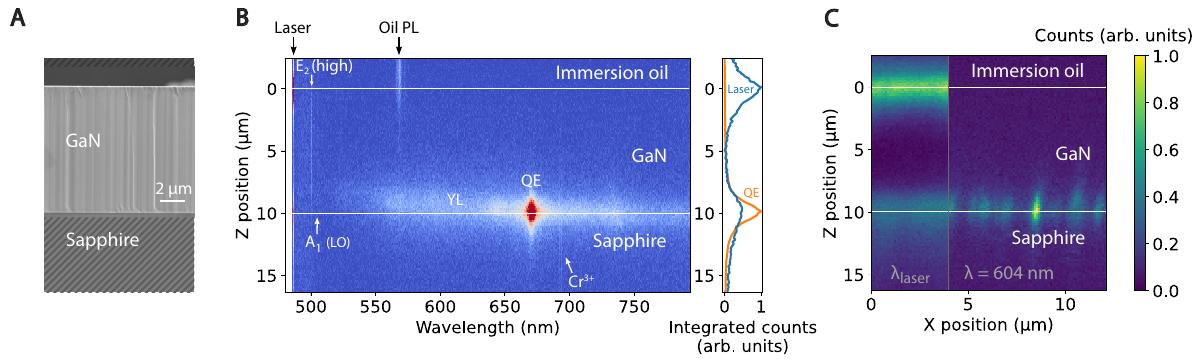}
\caption{Depth of naturally occurring emitters in GaN on sapphire. (A) Cross-sectional SEM image of a 10 µm-thick GaN layer. Scale bar: 2 µm. The hatched part of the image is digitally added to match the PL measurement scale. (B) Left: Depth-dependent PL spectrum of a QE embedded in the sample imaged in (A) and emitting at 671 nm. Right: Depth-dependent PL intensity averaged over the laser line (blue) and the QE ZPL (orange). (C) $XZ$ PL map of the same sample plotted at the laser wavelength to highlight the GaN interfaces (left) and at the emission wavelength of a QE (right).}\label{Fig_2}
\end{figure}

To probe the axial (vertical) distribution of emitters, we employ a high numerical aperture (NA) oil-immersion objective (NA$ = 1.45$), achieving a nearly diffraction-limited in-plane (lateral) resolution of 216 nm and an axial resolution of 1.29 µm (see SI Section S1). We intentionally investigate a 10 µm-thick GaN layer, ensuring that the sample thickness significantly exceeds the axial resolution of our objective. A cross-sectional scanning electron microscopy (SEM) image of the cleaved sample is displayed in Fig. \ref{Fig_2} (A). The sample is probed from the top with the µ-PL setup, and emitter depths are determined by performing a scan of the objective focal point along the GaN $c$-axis with a piezo stage ($Z$ scan) at the in-plane emitter positions. The emitter depths can then be compared to the GaN/oil and GaN/sapphire interfaces, identified through the laser reflection peaks. The $E_2$ (high) and $A_1$ (LO) Raman modes of GaN \cite{Raman_GaN_1995} provide an additional confirmation of the GaN epilayer position within the measurement, while the characteristic doublet around 694 nm associated with the Cr$^{3+}$ impurity present in sapphire indicates the sapphire substrate position. Given the refractive index mismatch between the immersion oil ($n_{\text{oil}}(\text{488 nm}) \simeq 1.52$) and GaN ($n_{\text{GaN}}(\text{488 nm}) \simeq 2.45$) a correction factor has to be applied to the depth scale (more information is given in the SI Section S1). The defect under investigation in Fig. \ref{Fig_2} (B), emitting at 671 nm, is found near the GaN/sapphire interface. This position coincides with the region of the broadband GaN YL\cite{reshchikov_origin_2023} which is expected to appear within the first 300 nm of the GaN epilayer \cite{tu_yellow_1998}. 

To verify whether all the emitters are located near the GaN/sapphire interface, we acquire a PL map by scanning the focal point of the objective both laterally and axially across the sample ($XZ$ PL map). At each point, a PL spectrum is recorded, allowing us to plot the map at different wavelengths. Figure \ref{Fig_2} (C) presents a single map plotted at the laser wavelength for the left-hand side of the map and at the emission wavelength of an emitter (604 nm, 0.3 nm bandwidth) on the right-hand side. The left panel allows us to identify the oil/GaN (top) and GaN/sapphire (bottom) interfaces where the laser is reflected to a greater extent. The right-hand side reveals the emitters with a ZPL or a phonon sideband contribution close to the plotted wavelength. All imaged QEs systematically align with the GaN/sapphire interface, independent of the chosen plotting wavelength. Additionally, in-plane maps acquired away from the GaN/sapphire interface show no localized emitters: an example taken 4 µm above the interface is provided in Fig. S2 (SI). These results demonstrate that emitters are not randomly scattered throughout the GaN thin film. They are localized close to the interface, which greatly hinders the potential integration of as-grown QEs within photonic structures such as $\lambda$-cavity micropillars,\cite{somaschi_near-optimal_2016} circular Bragg gratings,\cite{meunier_2023} and waveguides.\cite{lu_bright_2020} 

\begin{figure}[h!]
\centering
\includegraphics[width=1.\textwidth]{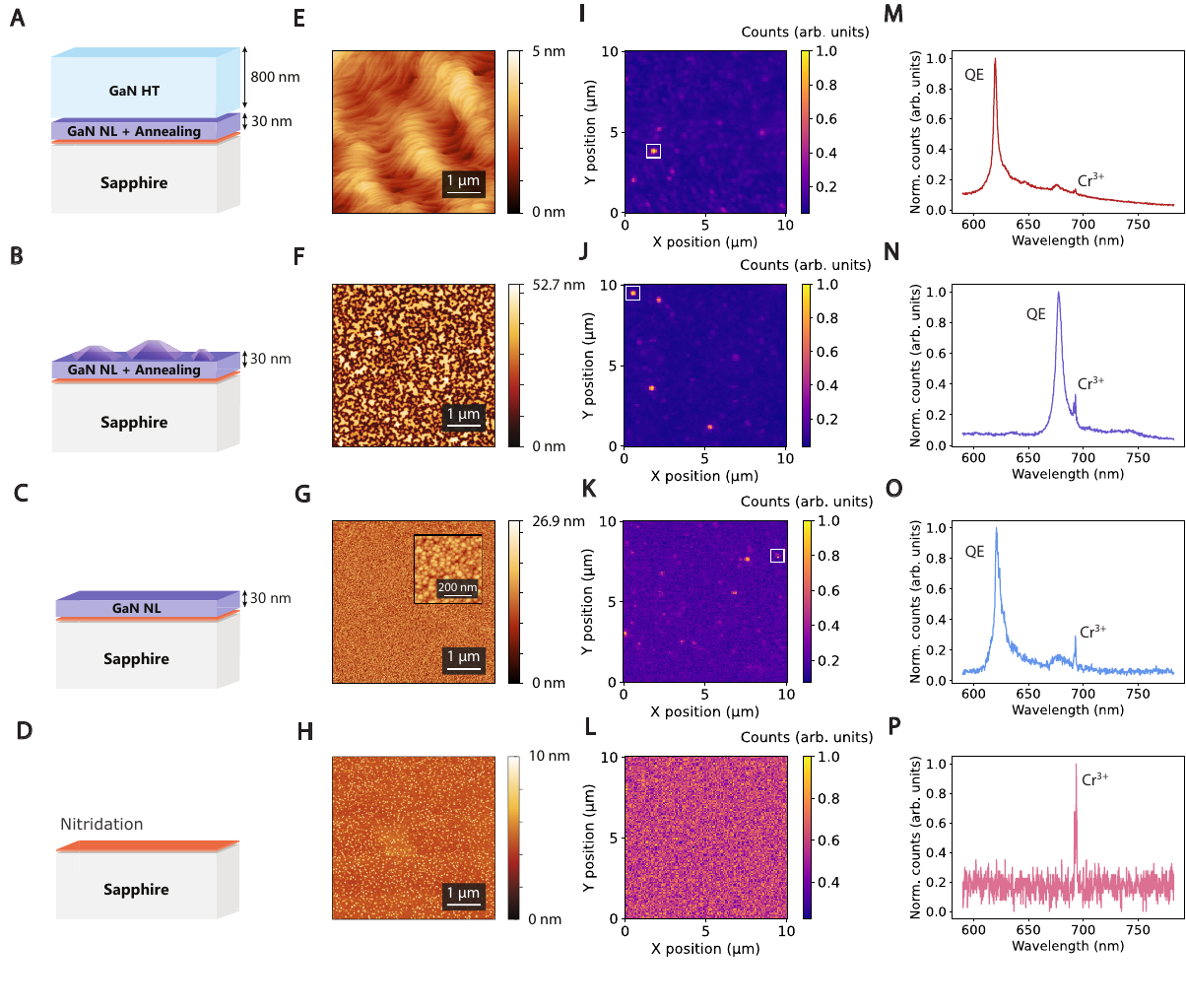}
\caption{Growth steps involved in the formation of naturally occurring emitters in GaN on sapphire. (A-D) Schematic of the investigated sample structures employed to identify the growth steps involved in QE formation. The depicted orange layer indicates that the sapphire has undergone nitridation. (E-H) AFM scans of the sample
surfaces. (I-K) PL maps at the emission wavelengths of the boxed emitters (0.3 nm bandwidth). The QEs are still present in the nucleation layer of GaN on sapphire. (L) PL map of the nitridated sapphire sample. As no emitter was observed, the map is displayed at 640 nm, chosen as a representative wavelength. (M-O) PL spectra of the highlighted emitters in the associated maps. (P) Typical PL spectrum of the nitridated sapphire substrate.} \label{Fig_3}
\end{figure}

To gain deeper insight into the formation of QEs in GaN, we focus on identifying the growth step resulting in their introduction. Indeed, the MOVPE growth of GaN on sapphire substrates involves several steps. First, the sapphire undergoes a nitridation process, which consists in exposing the surface to an ammonia (NH$_3$) flow at 1130 °C. This allows to control the polarity of the subsequent GaN layer (Figs. \ref{Fig_3} (D), (H)). Next, a thin GaN layer, approximately 30 nm thick, the so-called nucleation layer (NL), is deposited at low temperature (LT - 530 °C) using trimethylgallium (TMGa) and NH$_3$ as precursors (Figs. \ref{Fig_3} (C), (G)). This layer is then annealed at 1000 °C, leading to the formation of 50 nm-high islands (Figs. \ref{Fig_3} (B), (F)), which play a critical role in reducing the threading dislocation density \cite{nakamura_gan_NL_1991}. Finally, the GaN layer is grown to the desired thickness at high temperature (HT) $\sim$1000 °C (Figs. \ref{Fig_3} (A), (E)).
The investigated structures are shown in Figs. \ref{Fig_3} (A)-(D), with their corresponding surface morphologies characterized by atomic force microscopy (AFM) in Figs. \ref{Fig_3} (E)-(H). 
We first perform µ-PL measurements on a standard, coalesced, 800 nm thick GaN-on-sapphire template, where QEs are expected and indeed detected (Figs. \ref{Fig_3} (I), (M)). 
We then simplify the sample structures step by step to pinpoint the origin of the emitters. In the annealed NL, QEs are still measured with a comparable density and brightness to the GaN template (Figs. \ref{Fig_3} (J), (N)). In the as-deposited NL, we detect emitters as well, but with a significantly lower brightness (Figs. \ref{Fig_3} (K), (O)). Finally, we do not detect any emitters in the nitridated sapphire sample (Figs. \ref{Fig_3} (L), (P)). These measurements show that the defects responsible for quantum emission are formed in the LT GaN layer and become more optically active following the subsequent annealing step.

\subsection{Growth control of QEs in GaN}

\begin{figure}[h!]
\centering
\includegraphics[width=.71\textwidth]{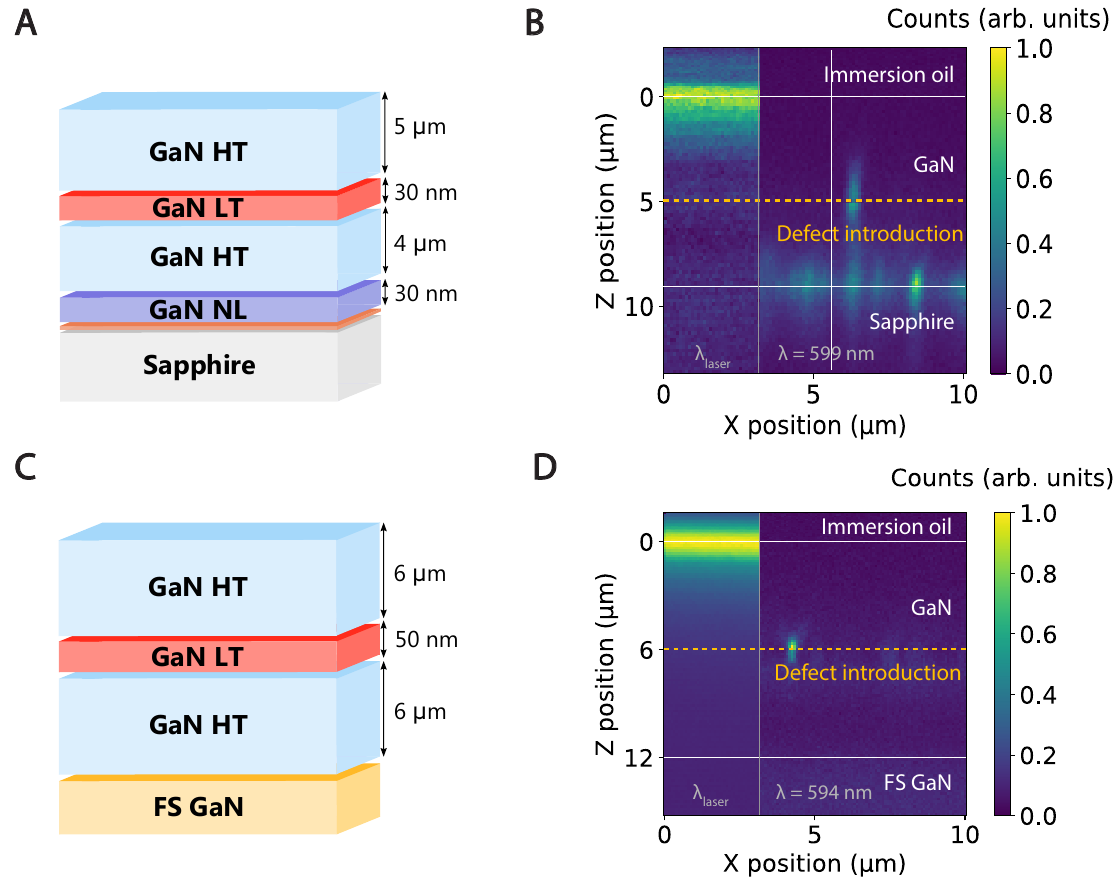}
\caption{Controlled depth positioning of QEs in GaN. (A) Structure of the sample grown on sapphire, designed to introduce QEs in GaN at a controlled depth. (B) $XZ$ PL map of the sample shown in (A), revealing naturally occurring emitters at the GaN/sapphire interface and an intentionally introduced emitter at the depth of the LT layer. The left-hand side is plotted at the laser wavelength to highlight the GaN interfaces, while the right-hand side is plotted at the emission wavelength of the introduced emitter located 5 µm within the layer ($\lambda=599$ nm). (C) Structure of the sample grown on FS GaN. (D) $XZ$ PL map of the sample depicted in (C), plotted at the laser wavelength and at the emission wavelength of an emitter introduced 6 µm within the layer ($\lambda=594$ nm).}\label{Fig_4}
\end{figure}

Next, we apply this finding to engineer emitters and introduce them at defined depths within the GaN layer, a key requirement for efficient emitter-cavity coupling. To this end, we prepare a sample consisting in a 4 µm-thick GaN-on-sapphire template on which, after a growth interruption to stabilize the surface temperature around 530 °C, 30 nm of LT GaN is deposited, followed by another growth interruption allowing the deposition of a 5 µm-thick HT GaN capping layer (Fig. \ref{Fig_4} (A)). With this design, we expect QEs both at the GaN/sapphire interface and 5 µm below the GaN surface. These distances are chosen so that they can be discriminated given the axial resolution of the setup. Figure \ref{Fig_4} (B) presents an $XZ$ PL map of the modified GaN sample, where an emitter has been successfully introduced at the depth of the LT GaN layer. In-plane PL maps acquired with the focal plane positioned at the depth of this LT layer (Fig. S2 (C) of the SI) further reveal a density of generated QEs of $\sim 2\times 10^7  \text{ cm}^{-2}$, about an order of magnitude lower than the density measured at the GaN/sapphire interface ($\sim 3\times 10^8 \text{ cm}^{-2}$), when considering the 520-780 nm wavelength range. This density remains sufficiently high to find multiple emitters within a single PL map, while still allowing individual QEs to be spectrally and spatially isolated. The optical properties of intentionally introduced emitters resemble greatly the properties of emitters found in the NL, with sharp ZPLs spanning the visible and the NIR (530-920 nm, mean FWHM: 5.4 nm), high Debye-Waller factors ranging from 0.69 to 0.98 (mean: 0.81), linearly polarized and bright antibunched emission (see Fig. S4 in the SI section S4). Notably, these figures of merit exceed those reported for silane-based approaches, where QEs exhibit substantially broader ZPLs and lower Debye-Waller factors \cite{eggleton_controlled_2026}. SIMS measurements were performed to link emitter presence with impurity concentrations (see Fig. S3 in the SI Section S3). These measurements reveal carbon, oxygen and hydrogen concentrations larger than $10^{19} \text{ cm}^{-3}$ both in the LT GaN layer and at the GaN/sapphire interface. No silicon concentration spike is detected near the emitter-forming layer, reducing parasitic doping and thus spectral jittering.\cite{kuhlmann_charge_2013} Interestingly, the oxygen concentration is lower by one order of magnitude in the LT GaN layer with respect to the NL, thus correlating with the defect densities, although not with a one to one ratio. This result suggests that oxygen may be part of the defect complex(es) acting as QE(s) in GaN. However, further experimental and theoretical work is required to validate this hypothesis regarding the nature of the defect(s).

To verify whether the QE introduction is independent on the substrate nature, we use the same growth process to generate QEs in a GaN thin film grown on free-standing (FS) GaN (Fig. \ref{Fig_4} (C)). The successful introduction of QEs in such a layer is demonstrated in Fig. \ref{Fig_4} (D) using an $XZ$ PL map, where an emitter is detected at the depth of the LT GaN layer. This result shows that defect introduction is substrate-independent and does not result exclusively from the migration of substrate impurities into the GaN layer. Furthermore, QEs had never been reported in GaN thin films grown on FS GaN, likely because the growth does not necessitate a LT GaN layer. The capability to introduce QEs in such layers is of great technological importance for the GaN quantum emission platform since excellent  distributed Bragg reflectors (DBRs) can be epitaxially grown on FS GaN \cite{cosendey_blue_2012, kobayashi_dbr_2024}, owing to its very low density of threading dislocations ($\sim 10^6 \text{ cm}^{-2}$). The capability to introduce QEs both in GaN layers grown on sapphire and on FS GaN thus enables the fabrication of high quality factor optical cavities embedding GaN QEs precisely positioned in depth. Finally, the substrate-independence of our approach for introducing emitters in GaN, together with the absence of any requirement for silane treatment, strongly suggests that in silane-based methods\cite{eggleton_controlled_2026} the key step responsible for emitter formation in GaN on silicon is not the silane treatment itself, but rather the LT GaN layer grown prior to subsequent high-temperature capping.

\subsection{Layer quality and emitter localization}

\begin{figure}[h!]
\centering
\includegraphics[width=1.\textwidth]{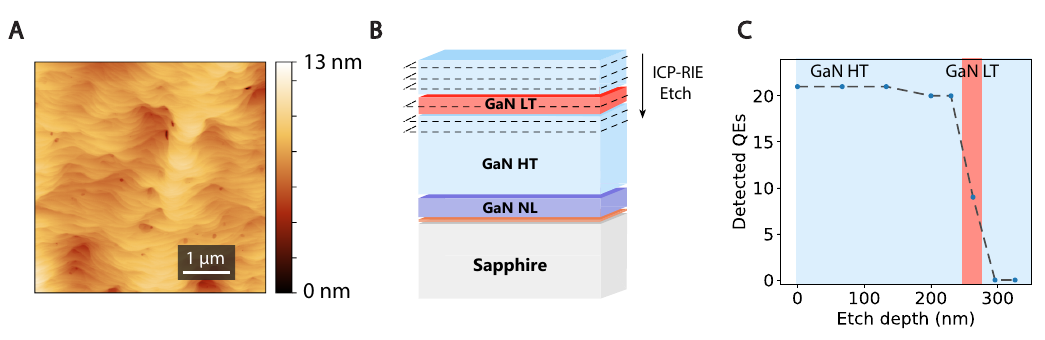}
\caption{Localization of intentionally introduced emitters within the GaN layer. (A) 5 µm $\times$ 5 µm AFM scan of a sample in which emitters are purposely formed 250 nm below the surface, prior to any fabrication step. The surface is fully coalesced, with an RMS surface roughness of 0.94 nm. (B) Schematic of the sample structure and measurement sequence, with successive µ-PL measurements and ICP-RIE etching. (C) Number of detected QEs within the same sample area as a function of cumulative etch depth. The QEs disappear at the depth of the GaN LT layer. }\label{Fig_5}
\end{figure}

Finally, we confirm that the introduced emitters lie within the LT GaN layer, thus demonstrating a vertical positioning accuracy below 60 nm. To this end, we grow a modified 6 µm-thick HT GaN template on sapphire incorporating a 30 nm-thick LT GaN interlayer positioned 250 nm below the surface. Full coalescence of the GaN film after the emitter-forming layer and the 250 nm cap is confirmed by AFM (Fig. \ref{Fig_5} (A)), yielding a low root mean square (RMS) surface roughness of 0.94 nm over a 5 µm $\times$ 5 µm scan area. Given the impurity-depth profile shown in Fig. S3 (B) of the SI Section S3, coalescence is expected to occur within less than $\sim$60 nm. Achieving a flat surface is essential for positioning doped layers or mirrors at well-defined distances from the QEs, enabling the engineering of quantum photonic devices. Notably, such full coalescence was not achieved in the recent report relying on silane treatment, which required a cap layer thicker than 3 µm.\cite{eggleton_controlled_2026}

To determine the depth of the introduced emitters, we first fabricate SiO$_2$ alignment markers on the sample surface to enable repeated µ-PL measurements at the same location. An initial µ-PL map is acquired with the focal plane positioned just below the surface. The sample is then subjected to successive inductively coupled plasma reactive-ion etching (ICP-RIE) steps, each followed by a µ-PL map of the same region using the alignment markers for accurate repositioning. Etch increments of 70 nm are used while far from the LT layer and reduced to 30 nm when approaching it to improve depth resolution (Fig. \ref{Fig_5} (B)). The depth of the LT GaN layer was determined independently from cross-sectional SEM measurements. The results, shown in Fig. \ref{Fig_5} (C), reveal that the emitters disappear only after the LT GaN layer has been completely removed, confirming that they are localized within the LT GaN layer or in its immediate vicinity. No emitters are detected in the HT GaN layer (see SI Sections S2 and S5), placing an upper limit on the emitter density in the HT GaN that is three orders of magnitude lower than in the LT GaN layer. After the full etching sequence, the RMS surface roughness is only 4.2 nm over the probed area, indicating that the emitter depth is determined with an accuracy of around 60 nm. This value sets an upper bound on the precision of emitter positioning using our growth strategy. Such a vertical accuracy is a critical milestone to enhance emitter-cavity coupling in future defect-based GaN quantum light sources. Additionally, capacitance-voltage (C-V) profiling was used to establish an upper limit of $1\times 10^{16}$ cm$^{-3}$ for the net doping concentration of the LT GaN interlayer (see Methods), well below typical intentional doping levels. Despite its high impurity concentration, the layer is therefore highly compensated. This low net doping is advantageous for device integration, as the LT GaN interlayer can effectively introduce QEs and simultaneously act as an intrinsic region when embedded within a p-i-n diode structure. 

\section{Conclusions}

In this study, we demonstrate the controlled introduction of radiative defects acting as QEs in GaN epilayers at predefined depths using a thin LT GaN interlayer, achieving an in-depth positioning accuracy better than 60 nm. These emitters are promising candidates for RT quantum emission since they can exhibit high count rates ($>$1.5 MHz) and strong antibunching ($g^{(2)}(0)=0.06$). In GaN on sapphire layers, we show that they are formed during the nucleation layer deposition and that their original location thus lies in the vicinity of the sapphire substrate. Building on this insight, we inserted a LT GaN layer between HT GaN layers and successfully introduced QEs at chosen depths within GaN films. We demonstrate that this approach is substrate-independent by generating QEs in GaN layers grown both on sapphire and FS GaN substrates. Our method preserves unintentionally doped layers and enables full coalescence within 250 nm. Combined with the high-quality waveguides and cavities already available on the GaN platform \cite{houdre_phc_2012, aharonovich_microdisk_2014}, this capability opens a clear path toward embedding GaN QEs in photonic structures with optimal spatial emitter-structure alignment. The low net doping of the LT GaN layer also enables QE integration into p-i-n architectures for reduced spectral diffusion\cite{anderson_electrical_2019} and electrical tuning\cite{somaschi_near-optimal_2016} of their emission wavelength. Altogether, these advances establish a solid foundation for the development of an efficient and tunable GaN-based quantum light source operating at room temperature.

\section{Experimental section}

\subsection{Sample Growth}

The growth of the samples was carried out in a horizontal MOVPE reactor (Aixtron 200/4 RF-S) on commercial $c$-plane sapphire and FS GaN substrates. All the studied samples were grown using TMGa and NH$_3$ as precursors, with H$_2$ as the carrier gas. For GaN growth on sapphire, typical GaN templates involved multiple growth steps: the substrate first underwent a nitridation step at 1130 °C under a flow of NH$_3$ and H$_2$. A NL was then deposited at 530 °C with a V/III ratio of 1400 and a growth rate of 300 nm.h$^{-1}$. The NL was subsequently annealed at 1000 °C under NH$_3$ flow and with H$_2$ as the carrier gas. HT GaN layers were then deposited at the same temperature with a V/III ratio of 500 and a growth rate of 2.3 µm.h$^{-1}$. In the case of the LT layers used to introduce emitters at a controlled depth, the growth was performed in the same conditions as the NL. The temperature was then ramped up to 1000 °C during a 5 minute growth interruption, providing in situ annealing under NH$_3$ flow with H$_2$ as the carrier gas, before subsequent HT GaN capping. Concerning the growth on FS GaN substrates, the latter exhibit a low threading dislocation density of approximately 10$^6 \text{ cm}^{-2}$, and the growth was directly initiated at high temperature.

\subsection{Experimental setup}

All µ-PL measurements were performed at room temperature. A CW 488 nm monomode laser is focused onto the sample using an oil-immersion objective (NA = 1.45, Nikon) to selectively address individual QEs. The objective is mounted on an $XYZ$ nanopositioning stage (PI P-611.3S) to acquire in-plane and $XZ$ PL maps of the sample. The emitted signal is collected through the same objective and directed either to a spectrometer equipped with a Si-based CCD for spectral analysis or to single-photon avalanche photodiodes (Excelitas) for photon counting. For \textit{g}$^{(2)}(\tau)$ measurements, the collected signal is equally split by a free-space 50:50 beamsplitter and detected by the avalanche photodiodes, which are connected to a time-tagging device (quTAG from qutools).

\subsection{Etching of the layer}

Before the successive etching steps, 200 nm-thick SiO$_2$ markers were fabricated on the sample by photolithography and lift-off using the positive photoresist AZ ECI 3007 together with the lift-off resist LOR 5A. These markers served both as alignment references for the precise repositioning of the µ-PL maps after each etching step and as reference features for accurate etch-depth determination owing to their high etch selectivity (15:1) relative to GaN. A reference Si sample coated with SiO$_2$ was etched simultaneously to monitor the SiO$_2$ etch depth, which was then used to determine the corresponding GaN etch depth. The exposed GaN was etched by ICP-RIE using a Cl$_2$/Ar plasma. The etched depth and resulting roughness were determined on both samples after each etching step by 3D optical profilometry measurements using the Sensofar S-Neox. 

\subsection{Ridge Fabrication}

The fabrication of the ridges containing the low temperature GaN layer was done in a single photolithography step. After the deposition, exposure, and development of the positive photoresist AZ 3027, ICP-RIE using Cl$_2$ and Ar was employed to etch 465 nm of the exposed GaN. The LT GaN layer being positioned 250 nm below the surface, this process allowed to completely etch away the LT GaN layer outside of the ridges. The etched depth was determined to be 465 nm, both by AFM and profilometry measurements.

\subsection{Capacitance-voltage profiling}

The measurements were performed using a wafer profiler CVP21 with a 0.01 M aqueous KOH solution as electrolyte. To estimate the net doping concentration $\lvert N_D-N_A \rvert$
of the LT GaN interlayer, we grew a sample with the following layer sequence: sapphire/1 µm Si-doped n-GaN ($6\times 10^{18}$ cm$^{-3}$)/1 µm unintentionally doped HT GaN/250 nm LT GaN/100 nm HT GaN cap. The layers above the n-GaN remained fully depleted even under a 1 V forward bias. Assuming an abrupt junction, using the residual net doping concentration measured on a reference HT GaN sample ($5\times 10^{14}$ cm$^{-3}$), and treating the LT layer concentration as the only free parameter, solving Poisson's equation yields a maximum net doping concentration of $1\times 10^{16}$ cm$^{-3}$ for the LT layer that would be consistent with full depletion at 1 V forward bias. This value is an upper bound and sits well below typical intentional net doping concentrations.

\begin{acknowledgement}

\subsection*{Funding Source}
This work was supported by the Swiss National Science Foundation (SNSF) under grant no. 200020-215633.
\end{acknowledgement}

\begin{suppinfo}

Characterization of the setup resolution and details related to axial elongation. Additional in-plane photoluminescence maps. SIMS measurements of a GaN epilayer with intentionally introduced QEs. Additional optical characterization of intentionally introduced QEs. Additional data on the absence of QEs in the HT GaN.

\end{suppinfo}

\bibliography{Depth_control}

@article{Kimble2008,
  title = {The quantum internet},
  author = {Kimble, H. J.},
  journal = {Nature},
  volume = {453},
  number = {7198},
  pages = {1023--1030},
  year = {2008},
  doi = {10.1038/nature07127}
}

@article{Rastelli_gaas_qds,
  title = {Hierarchical {Self}-{Assembly} of $\mathrm{GaAs}/\mathrm{AlGaAs}$ {Quantum} {Dots}},
  author = {Rastelli, A. and Stufler, S. and Schliwa, A. and Songmuang, R. and Manzano, C. and Costantini, G. and Kern, K. and Zrenner, A. and Bimberg, D. and Schmidt, O. G.},
  journal = {Phys. Rev. Lett.},
  volume = {92},
  issue = {16},
  pages = {166104},
  numpages = {4},
  year = {2004},
  month = {Apr},
  publisher = {American Physical Society},
  doi = {10.1103/PhysRevLett.92.166104}
}

@article{schweickert_2017,
    author = {Schweickert, Lucas and Jöns, Klaus D. and Zeuner, Katharina D. and Covre da Silva, Saimon Filipe and Huang, Huiying and Lettner, Thomas and Reindl, Marcus and Zichi, Julien and Trotta, Rinaldo and Rastelli, Armando and Zwiller, Val},
    title = {On-demand generation of background-free single photons from a solid-state source},
    journal = {Applied Physics Letters},
    volume = {112},
    number = {9},
    pages = {093106},
    year = {2018},
    month = {02},
    issn = {0003-6951},
    doi = {10.1063/1.5020038}
}

@article{Quantum_boxes_marzin_1994,
  author = {Marzin, J. -Y. and G\'erard, J. -M. and Izra\"el, A. and Barrier, D. and Bastard, G.},
  title = {Photoluminescence of {Single} {InAs} {Quantum} {Dots} {Obtained} by {Self}-{Organized} {Growth} on {GaAs}},
  journal = {Phys. Rev. Lett.},
  volume = {73},
  issue = {5},
  pages = {716--719},
  numpages = {0},
  year = {1994},
  month = {Aug},
  publisher = {American Physical Society},
  doi = {10.1103/PhysRevLett.73.716}
}

@article{holmes_gan_qds,
author = {Holmes, Mark J. and Choi, Kihyun and Kako, Satoshi and Arita, Munetaka and Arakawa, Yasuhiko},
title = {{Room}-{Temperature} {Triggered} {Single} {Photon} {Emission} from a {III}-{Nitride} {Site}-{Controlled} {Nanowire} {Quantum} {Dot}},
journal = {Nano Letters},
volume = {14},
number = {2},
pages = {982-986},
year = {2014},
doi = {10.1021/nl404400d}}

@article{Sebastian_GaN_QD,
author = {Tamariz, Sebastian and Callsen, Gordon and Stachurski, Johann and Shojiki, Kanako and Butt{\'e}, Raphaël and Grandjean, Nicolas},
title = {Toward {Bright} and {Pure} {Single} {Photon} {Emitters} at 300 {K} {Based} on {GaN} {Quantum} {Dots} on {Silicon}},
journal = {ACS Photonics},
volume = {7},
number = {6},
pages = {1515-1522},
year = {2020},
doi = {10.1021/acsphotonics.0c00310},
}

@article{Kurtsiefer2000,
  title = {Stable {Solid}-{State} {Source} of {Single} {Photons}},
  author = {Kurtsiefer, C. and Mayer, S. and Zarda, P. and Weinfurter, H.},
  journal = {Physical Review Letters},
  volume = {85},
  number = {2},
  pages = {290--293},
  year = {2000},
  doi = {10.1103/PhysRevLett.85.290}
}

@article{luo_odmr_2024,
	title = {Room temperature optically detected magnetic resonance of single spins in {GaN}},
	language = {en},
	journal = {Nature Materials},
	author = {Luo, Jialun and Geng, Yifei and Rana, Farhan and Fuchs, Gregory D.},
	year = {2024},
	volume = {23},
    number = {4},
    pages = {512--518}, 
    doi = {10.1038/s41563-024-01803-5}
}

@article{Senellart2017,
  title = {High-performance semiconductor quantum-dot single-photon sources},
  author = {Senellart, Pascale and Solomon, Glenn and White, Andrew},
  journal = {Nature Nanotechnology},
  volume = {12},
  number = {11},
  pages = {1026--1039},
  year = {2017},
  doi = {10.1038/nnano.2017.218}
}

@article{zhou_2018,
author = {Yu Zhou  and Ziyu Wang  and Abdullah Rasmita  and Sejeong Kim  and Amanuel Berhane  and Zolt√°n Bodrog  and Giorgio Adamo  and Adam Gali  and Igor Aharonovich  and Wei-bo Gao },
title = {Room temperature solid-state quantum emitters in the telecom range},
journal = {Science Advances},
volume = {4},
number = {3},
pages = {eaar3580},
year = {2018},
doi = {10.1126/sciadv.aar3580}}

@article{Nakamura1992,
  title = {Hole compensation mechanism of p-type {GaN} films},
  author = {Nakamura, Shuji and Mukai, Takashi and Senoh, Masayuki},
  journal = {Japanese Journal of Applied Physics},
  volume = {31},
  number = {Part 2, No. 2A},
  pages = {L139--L142},
  year = {1992},
  doi = {10.1143/JJAP.31.L139}
}

@article{Castelletto2014,
  title = {A silicon carbide room-temperature single-photon source},
  author = {Castelletto, S. and Johnson, B. C. and Ivády, V. and Stavrias, N. and Umeda, T. and Gali, A. and Ohshima, T.},
  journal = {Nature Materials},
  volume = {13},
  pages = {151--156},
  year = {2014},
  doi = {10.1038/nmat3806}
}

@article{grosso_tunable_2017,
	title = {Tunable and high-purity room temperature single-photon emission from atomic defects in hexagonal boron nitride},
	volume = {8},
	issn = {2041-1723},
	doi = {10.1038/s41467-017-00810-2},
	number = {1},
	journal = {Nature Communications},
	author = {Grosso, Gabriele and Moon, Hyowon and Lienhard, Benjamin and Ali, Sajid and Efetov, Dmitri K. and Furchi, Marco M. and Jarillo-Herrero, Pablo and Ford, Michael J. and Aharonovich, Igor and Englund, Dirk},
	month = sep,
	year = {2017},
	pages = {705},
}

@article{chatterjee_room-temperature_2025,
	title = {Room-temperature high-purity single-photon emission from carbon-doped boron nitride thin films},
	language = {en},
	journal = {Science Advances},
	author = {Chatterjee, Arka and Biswas, Abhijit and Fuhr, Addis S and Terlier, Tanguy and Sumpter, Bobby G and Ajayan, Pulickel M and Aharonovich, Igor and Huang, Shengxi},
    volume = {11},
    number = {25},
    pages = {eadv2899},
	year = {2025},
    doi = {10.1126/sciadv.adv2899},
}

@article{Berhane2017,
  title = {Bright room-temperature single-photon emission from defects in {Gallium} {Nitride}},
  author = {Berhane, A. M. and Mendelson, N. and Nguyen, M. and Kim, S. and Rahman, A. and Tran, T. T. and Kianinia, M. and Toth, M. and Aharonovich, I.},
  journal = {Advanced Materials},
  volume = {29},
  number = {39},
  pages = {1605092},
  year = {2017},
  doi = {10.1002/adma.201605092}
}

@article{houdre_phc_2012,
    author = {Vico Triviño, N. and Rossbach, G. and Dharanipathy, U. and Levrat, J. and Castiglia, A. and Carlin, J.-F. and Atlasov, K. A. and Butté, R. and Houdré, R. and Grandjean, N.},
    title = {High quality factor two dimensional {GaN} photonic crystal cavity membranes grown on silicon substrate},
    journal = {Applied Physics Letters},
    volume = {100},
    number = {7},
    pages = {071103},
    year = {2012},
    month = {02},
    issn = {0003-6951},
    doi = {10.1063/1.3684630}
}

@article{berhane_photophysics_2018,
	title = {Photophysics of {GaN} single-photon emitters in the visible spectral range},
	volume = {97},
	issn = {2469-9950, 2469-9969},
	doi = {10.1103/PhysRevB.97.165202},
	language = {en},
	number = {16},
	urldate = {2023-09-22},
	journal = {Physical Review B},
	author = {Berhane, Amanuel M. and Jeong, Kwang-Yong and Bradac, Carlo and Walsh, Michael and Englund, Dirk and Toth, Milos and Aharonovich, Igor},
	month = apr,
	year = {2018},
	pages = {165202},
}

@article{geng_optical_2023,
	title = {Optical {Dipole} {Structure} and {Orientation} of {GaN} {Defect} {Single}-{Photon} {Emitters}},
	volume = {10},
	issn = {2330-4022, 2330-4022},
	doi = {10.1021/acsphotonics.3c00917},
	language = {en},
	number = {10},
	urldate = {2024-02-12},
	journal = {ACS Photonics},
	author = {Geng, Yifei and Jena, Debdeep and Fuchs, Gregory D. and Zipfel, Warren R. and Rana, Farhan},
	month = oct,
	year = {2023},
	pages = {3723--3729},
	}

@article{
aharonovich_microdisk_2014,
author = {Alexander Woolf  and Tim Puchtler  and Igor Aharonovich  and Tongtong Zhu  and Nan Niu  and Danqing Wang  and Rachel Oliver  and Evelyn L. Hu },
title = {Distinctive signature of indium gallium nitride quantum dot lasing in microdisk cavities},
journal = {Proceedings of the National Academy of Sciences},
volume = {111},
number = {39},
pages = {14042-14046},
year = {2014},
doi = {10.1073/pnas.1415464111}
}

@article{nguyen_effects_2019,
	title = {Effects of microstructure and growth conditions on quantum emitters in gallium nitride},
	volume = {7},
	issn = {2166-532X},
	doi = {10.1063/1.5098794},
	language = {en},
	number = {8},
	urldate = {2024-02-27},
	journal = {APL Materials},
	author = {Nguyen, Minh and Zhu, Tongtong and Kianinia, Mehran and Massabuau, Fabien and Aharonovich, Igor and Toth, Milos and Oliver, Rachel and Bradac, Carlo},
	month = aug,
	year = {2019},
	pages = {081106},
}

@article{Raman_GaN_1995,
doi = {10.1088/0953-8984/7/10/002},
year = {1995},
month = {mar},
publisher = {},
volume = {7},
number = {10},
pages = {L129},
author = {T Azuhata and T Sota and K Suzuki and S Nakamura},
title = {Polarized {Raman} spectra in {GaN}},
journal = {Journal of Physics: Condensed Matter},
}

@article{meunier_2023,
	title = {Telecom single-photon emitters in {GaN} operating at room temperature: embedment into bullseye antennas},
	volume = {12},
	language = {en},
	number = {8},
	journal = {Nanophotonics},
	author = {Meunier, Max and Eng, John J. H. and Mu, Zhao and Chenot, Sebastien and Brändli, Virginie and De Mierry, Philippe and Gao, Weibo and Zúñiga-Pérez, Jesús},
	month = apr,
	year = {2023},
	pages = {1405--1419},
    doi = {10.1515/nanoph-2022-0659},


}

@article{purcell_1946,
	title = {Spontaneous {Emission} {Probabilities} at {Radio} {Frequencies}},
	volume = {69},
	doi = {10.1103/PhysRev.69.674.2},
	journal = {Physical Review},
	author = {Purcell, E. M.},
	year = {1946},
	pages = {681}
}

@article{gerard_1998,
	title = {Enhanced {Spontaneous} {Emission} by {Quantum} {Boxes} in a {Monolithic} {Optical} {Microcavity}},
	volume = {81},
	issn = {0031-9007, 1079-7114},
	doi = {10.1103/PhysRevLett.81.1110},
	language = {en},
	number = {5},
	urldate = {2025-10-08},
	journal = {Physical Review Letters},
	author = {Gérard, J.-M. and Sermage, B. and Gayral, B. and Legrand, B. and Costard, E. and Thierry-Mieg, V.},
	month = aug,
	year = {1998},
	pages = {1110--1113},
}

@article{kaupp_postfabrication_2025,
	title = {Post-fabrication tuning of circular {Bragg} grating resonators via atomic layer deposition},
	volume = {127},
	issn = {0003-6951, 1077-3118},
	doi = {10.1063/5.0287371},
	language = {en},
	number = {13},
	urldate = {2025-10-08},
	journal = {Applied Physics Letters},
	author = {Kaupp, Jochen and Reum, Yorick and Peniakov, Giora and Emmerling, Monika and Estevam, Sabrina and Kamp, Martin and Huber-Loyola, Tobias and Höfling, Sven and Pfenning, Andreas Theo},
	month = sep,
	year = {2025},
	pages = {133501},
}

@article{krieger_postfabrication_2024,
	title = {Postfabrication {Tuning} of {Circular} {Bragg} {Resonators} for {Enhanced} {Emitter}-{Cavity} {Coupling}},
	volume = {11},
	issn = {2330-4022, 2330-4022},
	doi = {10.1021/acsphotonics.3c01480},
	language = {en},
	number = {2},
	urldate = {2025-10-08},
	journal = {ACS Photonics},
	author = {Krieger, Tobias M. and Weidinger, Christian and Oberleitner, Thomas and Undeutsch, Gabriel and Rota, Michele B. and Tajik, Naser and Aigner, Maximilian and Buchinger, Quirin and Schimpf, Christian and Garcia, Ailton J. and Covre Da Silva, Saimon F. and Höfling, Sven and Huber-Loyola, Tobias and Trotta, Rinaldo and Rastelli, Armando},
	month = feb,
	year = {2024},
	pages = {596--603},
}

@article{yuan_first-principle_2023,
	title = {First-{Principle} {Prediction} of {Stress}-{Tunable} {Single}-{Photon} {Emitters} at {Telecommunication} {Band} from {Point} {Defects} in {GaN}},
	volume = {10},
	issn = {2304-6732},
	doi = {10.3390/photonics10050544},
	language = {en},
	number = {5},
	urldate = {2024-03-25},
	journal = {Photonics},
	author = {Yuan, Junxiao and Wang, Ke and Hou, Yidong and Chen, Feiliang and Li, Qian},
	month = may,
	year = {2023},
	pages = {544},
}

@article{dousse_controlled_2008,
	title = {Controlled {Light}-{Matter} {Coupling} for a {Single} {Quantum} {Dot} {Embedded} in a {Pillar} {Microcavity} {Using} {Far}-{Field} {Optical} {Lithography}},
	volume = {101},

	issn = {0031-9007, 1079-7114},

	doi = {10.1103/PhysRevLett.101.267404},
	language = {en},
	number = {26},
	urldate = {2025-10-08},
	journal = {Physical Review Letters},
	author = {Dousse, A. and Lanco, L. and Suffczyński, J. and Semenova, E. and Miard, A. and Lemaître, A. and Sagnes, I. and Roblin, C. and Bloch, J. and Senellart, P.},
	month = dec,
	year = {2008},
	pages = {267404},

}

@article{wijitpatima_bright_2024,
	title = {Bright {Electrically} {Contacted} {Circular} {Bragg} {Grating} {Resonators} with {Deterministically} {Integrated} {Quantum} {Dots}},
	volume = {18},
	issn = {1936-0851, 1936-086X},
	doi = {10.1021/acsnano.4c07820},
	language = {en},
	number = {46},
	urldate = {2025-04-01},
	journal = {ACS Nano},
	author = {Wijitpatima, Setthanat and Auler, Normen and Mudi, Priyabrata and Funk, Timon and Barua, Avijit and Shrestha, Binamra and Schall, Johannes and Limame, Imad and Rodt, Sven and Reuter, Dirk and Reitzenstein, Stephan},
	month = nov,
	year = {2024},
	pages = {31834--31845},
}

@article{descamps_2024,
    doi = {10.1088/1361-6528/ad5dbd},
    year = {2024},
    month = {jul},
    publisher = {IOP Publishing},
    volume = {35},
    number = {41},
    pages = {415703},
    author = {Descamps, Thomas and Bampis, Alexandros and Huet, Maximilien and Hammar, Mattias and Zwiller, Val},
    title = {Mapping and spectroscopy of telecom quantum emitters with confocal laser scanning microscopy},
    journal = {Nanotechnology}
}

@article{sapienza_2015,
	title = {Nanoscale optical positioning of single quantum dots for bright and pure single-photon emission},
	volume = {6},
	issn = {2041-1723},
	doi = {10.1038/ncomms8833},
	language = {en},
	number = {1},
	urldate = {2025-10-08},
	journal = {Nature Communications},
	author = {Sapienza, Luca and Davanço, Marcelo and Badolato, Antonio and Srinivasan, Kartik},
	month = jul,
	year = {2015},
	pages = {7833},
}

@article{fournier_position-controlled_2021,
	title = {Position-controlled quantum emitters with reproducible emission wavelength in hexagonal boron nitride},
	volume = {12},
	issn = {2041-1723},
	doi = {10.1038/s41467-021-24019-6},
	language = {en},
	number = {1},
	urldate = {2024-03-25},
	journal = {Nature Communications},
	author = {Fournier, Clarisse and Plaud, Alexandre and Roux, Sébastien and Pierret, Aurélie and Rosticher, Michael and Watanabe, Kenji and Taniguchi, Takashi and Buil, Stéphanie and Quélin, Xavier and Barjon, Julien and Hermier, Jean-Pierre and Delteil, Aymeric},
	month = jun,
	year = {2021},
	pages = {3779},
}

@article{chen_laser_2017,
	title = {Laser writing of coherent colour centres in diamond},
	volume = {11},
	issn = {1749-4885, 1749-4893},
	doi = {10.1038/nphoton.2016.234},
	language = {en},
	number = {2},
	urldate = {2024-04-24},
	journal = {Nature Photonics},
	author = {Chen, Yu-Chen and Salter, Patrick S. and Knauer, Sebastian and Weng, Laiyi and Frangeskou, Angelo C. and Stephen, Colin J. and Ishmael, Shazeaa N. and Dolan, Philip R. and Johnson, Sam and Green, Ben L. and Morley, Gavin W. and Newton, Mark E. and Rarity, John G. and Booth, Martin J. and Smith, Jason M.},
	month = feb,
	year = {2017},
	pages = {77--80},

}

@article{tu_yellow_1998,
	title = {Yellow luminescence depth profiling on {GaN} epifilms using reactive ion etching},
	volume = {73},
	issn = {0003-6951, 1077-3118},
	doi = {10.1063/1.122595},
	language = {en},
	number = {19},
	journal = {Applied Physics Letters},
	author = {Tu, L. W. and Lee, Y. C. and Chen, S. J. and Lo, I. and Stocker, D. and Schubert, E. F.},
	month = nov,
	year = {1998},
	pages = {2802--2804},
}

@article{akbari_tunable_hbn_2022,
	title = {Lifetime-{Limited} and {Tunable} {Quantum} {Light} {Emission} in h-{BN} via {Electric} {Field} {Modulation}},
	volume = {22},
	issn = {1530-6984, 1530-6992},
	doi = {10.1021/acs.nanolett.2c02163},
	language = {en},
	number = {19},
	urldate = {2024-02-19},
	journal = {Nano Letters},
	author = {Akbari, Hamidreza and Biswas, Souvik and Jha, Pankaj Kumar and Wong, Joeson and Vest, Benjamin and Atwater, Harry A.},
	month = oct,
	year = {2022},
	pages = {7798--7803},

}

@article{cosendey_blue_2012,
	title = {Blue monolithic {AlInN}-based vertical cavity surface emitting laser diode on free-standing {GaN} substrate},
	volume = {101},
	issn = {0003-6951, 1077-3118},
	doi = {10.1063/1.4757873},
	language = {en},
	number = {15},
	urldate = {2025-10-08},
	journal = {Applied Physics Letters},
	author = {Cosendey, Gatien and Castiglia, Antonino and Rossbach, Georg and Carlin, Jean-François and Grandjean, Nicolas},
	month = oct,
	year = {2012},
	pages = {151113},
}

@article{reshchikov_origin_2023,
	title = {On the {Origin} of the {Yellow} {Luminescence} {Band} in {GaN}},
	volume = {260},
	issn = {0370-1972, 1521-3951},

	doi = {10.1002/pssb.202200488},
	
	language = {en},
	number = {8},
	urldate = {2025-10-15},
	journal = {{Physica} {Status} {Solidi} (b)},
	author = {Reshchikov, Michael A.},
	month = aug,
	year = {2023},
	pages = {2200488},

}

@article{nakamura_gan_NL_1991,
	title = {{GaN} {Growth} {Using} {GaN} {Buffer} {Layer}},
	volume = {30},
	issn = {0021-4922, 1347-4065},
	doi = {10.1143/JJAP.30.L1705},
	number = {10A},
	journal = {Japanese Journal of Applied Physics},
	author = {Shuji Nakamura},
	month = oct,
	year = {1991},
	pages = {L1705},
}

@article{kobayashi_dbr_2024,
	title = {In {Situ} {Center} {Wavelength} {Control} of {AlInN}/{GaN} {Distributed} {Bragg} {Reflectors} with {In} {Situ} {Reflectivity} {Spectra} {Measurements}},
	volume = {261},
	issn = {0370-1972, 1521-3951},
	doi = {10.1002/pssb.202400010},
	language = {en},
	number = {11},
	urldate = {2025-10-21},
	journal = {{Physica} {Status} {Solidi} (b)},
	author = {Kobayashi, Kenta and Nishikawa, Taichi and Watanabe, Ruka and Takeuchi, Tetsuya and Kamiyama, Satoshi and Iwaya, Motoaki and Kamei, Toshihiro},
	month = nov,
	year = {2024},
	pages = {2400010},

}

@article{odmr_telecom_2025,
  title = {{Room}-{Temperature} {Optically} {Detected} {Magnetic} {Resonance} of {Telecom} {Single}-{Photon} {Emitters} in {GaN}},
  author = {Eng, John J. H. and Jiang, Zhengzhi and Meunier, Max and Rasmita, Abdullah and Zhang, Haoran and Yang, Yuzhe and Zhou, Feifei and Cai, Hongbing and Dong, Zhaogang and Z\'u\~niga-P\'erez, Jes\'us and Gao, Weibo},
  journal = {Phys. Rev. Lett.},
  volume = {134},
  issue = {8},
  pages = {083602},
  numpages = {6},
  year = {2025},
  month = {Feb},
  publisher = {American Physical Society},
  doi = {10.1103/PhysRevLett.134.083602}
}

@article{bishop_SIL_2022,
	title = {Enhanced light collection from a gallium nitride color center using a near index-matched solid immersion lens},
	volume = {120},
	doi = {10.1063/5.0085257},
	language = {en},
	number = {11},
	urldate = {2024-06-05},
	journal = {Applied Physics Letters},
	author = {Bishop, S. G. and Hadden, J. P. and Hekmati, R. and Cannon, J. K. and Langbein, W. W. and Bennett, A. J.},
	month = mar,
	year = {2022},
	pages = {114001},

}

@article{eggleton_controlled_2026,
	title = {Controlled epitaxy of room-temperature quantum emitters in gallium nitride},
	volume = {11},
	issn = {2378-0967},
	doi = {10.1063/5.0300338},
	language = {en},
	number = {1},
	journal = {APL Photonics},
	author = {Eggleton, Katie M. and Cannon, Joseph K. and Bishop, Sam G. and Hadden, John P. and Zhao, Chunyu and Kappers, Menno J. and Oliver, Rachel A. and Bennett, Anthony J.},
	month = jan,
	year = {2026},
	pages = {016103},
}

@article{somaschi_near-optimal_2016,
	title = {Near-optimal single-photon sources in the solid state},
	volume = {10},
	issn = {1749-4893},
	doi = {10.1038/nphoton.2016.23},
	number = {5},
	journal = {Nature Photonics},
	author = {Somaschi, N. and Giesz, V. and De Santis, L. and Loredo, J. C. and Almeida, M. P. and Hornecker, G. and Portalupi, S. L. and Grange, T. and Antón, C. and Demory, J. and Gómez, C. and Sagnes, I. and Lanzillotti-Kimura, N. D. and Lemaítre, A. and Auffeves, A. and White, A. G. and Lanco, L. and Senellart, P.},
	month = may,
	year = {2016},
	pages = {340--345},
}

@article{lu_bright_2020,
	title = {Bright {High}-{Purity} {Quantum} {Emitters} in {Aluminum} {Nitride} {Integrated} {Photonics}},
	volume = {7},
	issn = {2330-4022, 2330-4022},
	doi = {10.1021/acsphotonics.0c01259},
	language = {en},
	number = {10},
	urldate = {2024-02-19},
	journal = {ACS Photonics},
	author = {Lu, Tsung-Ju and Lienhard, Benjamin and Jeong, Kwang-Yong and Moon, Hyowon and Iranmanesh, Ava and Grosso, Gabriele and Englund, Dirk},
	month = oct,
	year = {2020},
	pages = {2650--2657},
}

@article{anderson_electrical_2019,
	title = {Electrical and optical control of single spins integrated in scalable semiconductor devices},
	volume = {366},
	issn = {0036-8075, 1095-9203},
	doi = {10.1126/science.aax9406},
	language = {en},
	number = {6470},
	urldate = {2026-04-22},
	journal = {Science},
	author = {Anderson, Christopher P. and Bourassa, Alexandre and Miao, Kevin C. and Wolfowicz, Gary and Mintun, Peter J. and Crook, Alexander L. and Abe, Hiroshi and Ul Hassan, Jawad and Son, Nguyen T. and Ohshima, Takeshi and Awschalom, David D.},
	month = dec,
	year = {2019},
	pages = {1225--1230},
}

@article{kuhlmann_charge_2013,
	title = {Charge noise and spin noise in a semiconductor quantum device},
	volume = {9},
	copyright = {http://www.springer.com/tdm},
	issn = {1745-2473, 1745-2481},
	doi = {10.1038/nphys2688},
	language = {en},
	number = {9},
	urldate = {2026-04-22},
	journal = {Nature Physics},
	author = {Kuhlmann, Andreas V. and Houel, Julien and Ludwig, Arne and Greuter, Lukas and Reuter, Dirk and Wieck, Andreas D. and Poggio, Martino and Warburton, Richard J.},
	month = sep,
	year = {2013},
	pages = {570--575},
}

@book{nano-optics, 
    place={Cambridge}, 
    title={Principles of Nano-Optics}, 
    publisher={Cambridge University Press, Cambridge}, 
    author={Novotny, Lukas and Hecht, Bert},
    year={2012}, 
    chapter = {"Resolution and Localization"}}

@article{hell_aberrations_1993,
	title = {{Aberrations} in confocal fluorescence microscopy induced by mismatches in refractive index},
	volume = {169},
	issn = {0022-2720, 1365-2818},
	language = {en},
	journal = {Journal of Microscopy},
	author = {Hell, S. and Reiner, G. and Cremer, C. and Stelzer, E. H. K.},
	month = mar,
	year = {1993},
	pages = {391--405},
}

@article{everall_confocal_2009,
	title = {Confocal {Raman} {Microscopy}: {Performance}, {Pitfalls}, and {Best} {Practice}: {Invited} {Lecture} at the {Symposium} “50 {Years} of {SAS}: {Looking} to the {Future} with {Vibrational} {Spectroscopy}” at {Pittcon} 2008, {New} {Orleans}, {Louisiana}},
	volume = {63},
	issn = {0003-7028, 1943-3530},
	shorttitle = {Confocal {Raman} {Microscopy}},
	language = {en},
	journal = {Applied Spectroscopy},
	author = {Everall, Neil J.},
	month = sep,
	year = {2009},
	pages = {245A--262A},
}

\end{document}


\title{Supporting Information for:\\Depth Control of Room-Temperature Quantum Emitters in Gallium Nitride}
\author{
    Alexandros Bampis$^{*}$, Johann Stachurski, Anna Schwab, \\ Jean-François Carlin, Raphaël Butté, and Nicolas Grandjean
}
\date{}
\maketitle

\begin{center}
Institute of Physics, Ecole Polytechnique Fédérale de Lausanne (EPFL),\\CH-1015 Lausanne, Switzerland\\
$^{*}$Corresponding author: \texttt{alexandros.bampis@epfl.ch}
\end{center}

\tableofcontents
\newpage

\section{Setup resolution and axial elongation}

According to Abbé's diffraction formulas, the lateral and axial resolution ($R_{\text{lat}}$ and $R_{\text{ax}}$, respectively) of the setup are expected to be:\cite{nano-optics}

\begin{equation}
    \left\{
    \begin{array}{ll}
        R_{\text{lat}} = \frac{0.61\lambda}{ \text{NA}} \simeq 205 \text{ nm}, \\
        R_{\text{ax}} = \frac{2n\lambda}{\text{NA}^2} \simeq 706 \text{ nm},
    \end{array}
\right.
\end{equation}
where $\lambda = 488$ nm is the excitation wavelength, $\text{NA} = 1.45$ is the numerical aperture of the oil-immersion objective and $n\simeq 1.52$ is the refractive index of the immersion oil. This optimal performance, however, can only be obtained when the studied sample has a refractive index close to that of the immersion oil (or  is immersed in such a medium). In our case, the refractive index of GaN ($n_{\text{GaN}} \simeq 2.45$  at $\lambda = 488\text{ nm}$) strongly differs from that of the immersion oil ($n_{\text{oil}} \simeq 1.52$ at $\lambda = 488\text{ nm}$). As a result, light rays are refracted at the oil/GaN interface, which stretches the focal spot, changing the depth scale and inducing spherical aberrations — arising from the off-axis distribution of the laser intensity at the interface.\cite{hell_aberrations_1993, everall_confocal_2009}

Since the numerical correction of these effects can be computationally demanding, we employ an empirical approach to correct the depth axes of both $XZ$ micro-photoluminescence (µ-PL) maps and $Z$ scans. This correction is achieved by matching the optical profiles to the layer thicknesses measured with cross-sectional scanning electron microscopy (SEM) imaging and secondary ion mass spectrometry (SIMS) depth profiling. The position of the oil/GaN interface in the PL measurements is given by the maximum laser reflection. For the GaN/substrate interface, the second laser reflection can be used as a reference to locate the sapphire substrate, while for free-standing GaN, where no reflection occurs, we instead use the background luminescence originating from the substrate, which is brighter than that of the epilayer.

Specifically, in our measurements, a displacement of the $Z$ piezo by 1 µm typically induces a shift in the focus spot depth within the GaN of around 1.7 µm. In all $Z$ and $XZ$ measurements, the correction of the depth scale is only applied after the oil/GaN interface. The achievable depth accuracy in our measurements then becomes $R_{\text{ax}}^{\text{corr}}\simeq 1.7 \times R_{\text{ax}} \simeq 1.2 \text{ µm}$.

\begin{figure}[h!]
\centering
\includegraphics[width=0.75\textwidth]{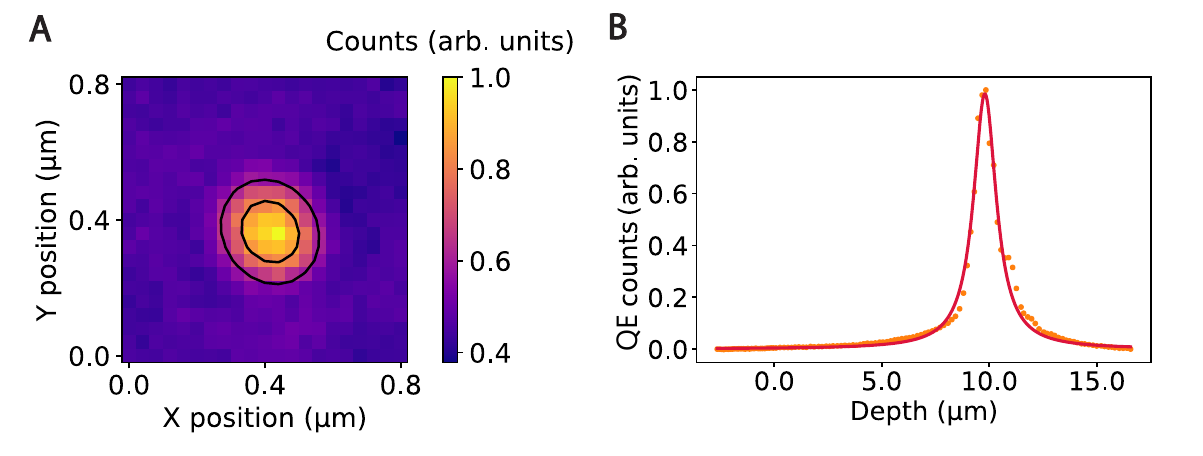}
\caption{(A) PL map centered on an isolated QE. The QE is fitted with a 2D Gaussian fit (FWHM$=216$ nm). The black lines represent the contours of the fit. (B) Depth-dependent counts integrated over the ZPL of a QE embedded in a 10 µm-thick GaN on sapphire layer. The depth profile is fitted with a Lorentzian function (red curve), yielding a FWHM of 1.29 µm.}\label{Fig_resolution}
\end{figure}

We first assess the lateral resolution of our setup by performing a high resolution PL map on a well-isolated quantum emitter (QE) (Fig. \ref{Fig_resolution} (A)). The QE profile is then fitted with a 2D Gaussian fit which yields a full width at half maximum (FWHM) corresponding to the lateral resolution of the setup since the QE is a point-like source. We thus have an estimate for the lateral resolution of the setup: $R_{\text{lat}}^{\text{exp}}=216$ nm. In Fig. \ref{Fig_resolution} (B), we assess the axial resolution of the setup by performing a $Z$ scan at the in-plane position of an emitter, which lies in the nucleation layer (NL) of the 10 µm-thick GaN on sapphire sample. The counts are integrated over the zero-phonon line (ZPL) of the emitter, and the profile is then fitted with a Lorentzian function yielding an estimate of the axial resolution: $R_{\text{ax}}^{\text{exp}} = 1.29 \text{ µm}$.
 
Both values are comparable with the expected resolutions. The resulting axial resolution is therefore sufficient to distinguish emitters located in the GaN NL from those introduced in the low-temperature (LT) GaN layers of the samples presented in the main text. We note that enhanced resolutions could be achieved using a confocal detection scheme. However, this comes at the cost of a heavy loss of signal, and we have chosen here to work without a confocal path to be able to perform large spatial PL maps.

\section{In-plane PL maps}

To assess the spatial distribution of emitters within the GaN layer, we performed µ-PL mapping on the 10 µm-thick GaN sample and on the sample containing intentionally introduced emitters presented in Fig. 4 (A) of the main text. In contrast to the map presented in Fig. 1 (A) of the main text, these maps display the maximum intensity over the spectral range of the measurement (480-780 nm) at each spatial position, while excluding contributions from the excitation laser, Raman peaks, oil photoluminescence, and the Cr$^{3+}$ emission lines of sapphire. This approach enables a comprehensive visualization of all emitters present within the probed spectral window. In the following, these PL maps will be referred to as panchromatic maps.

\begin{figure}[h!]
\centering
\includegraphics[width=0.76\textwidth]{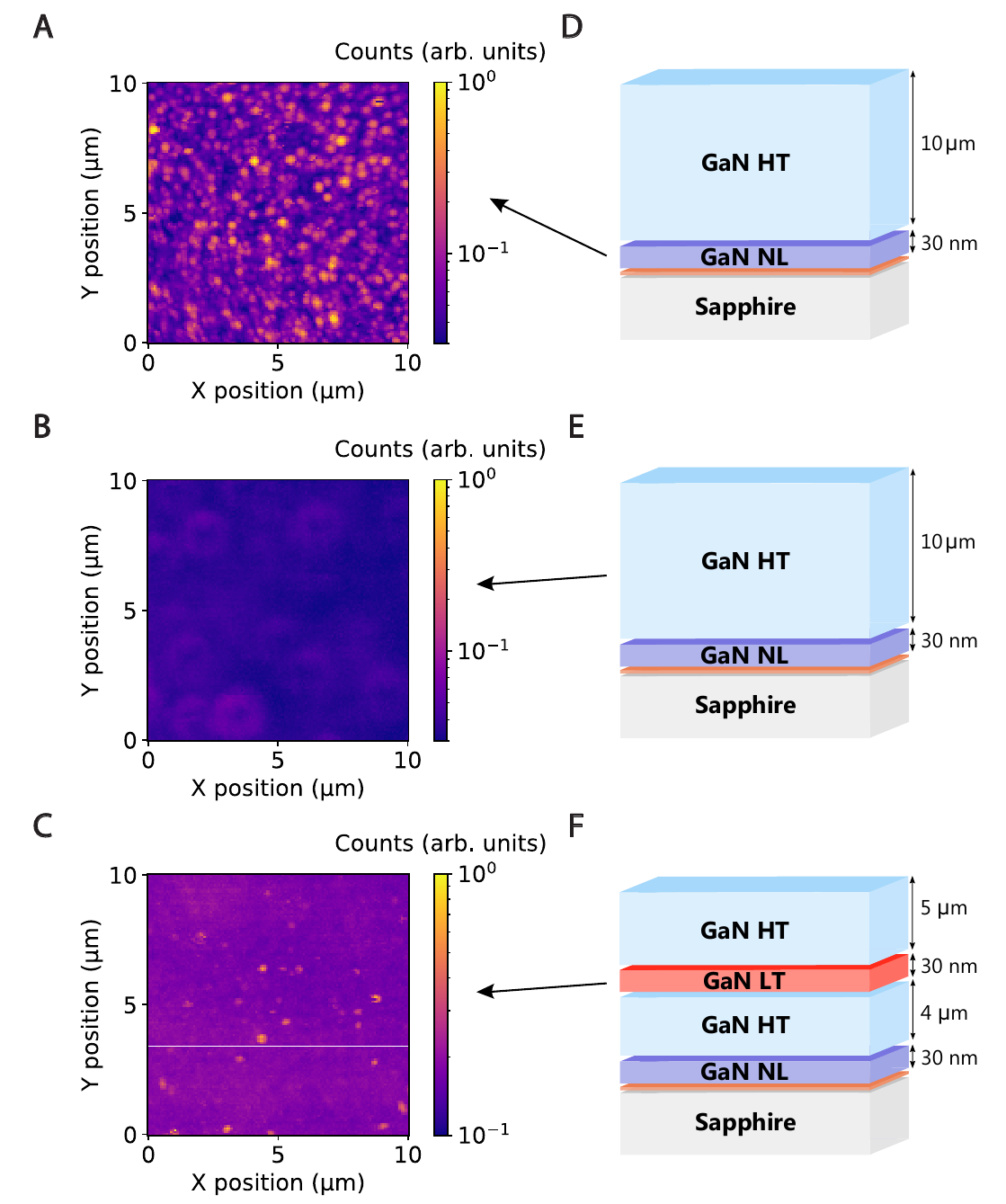}
\caption{(A) Panchromatic PL map acquired at the depth of the NL, revealing a high density of localized emitters over the 520-780 nm wavelength range.
(B) Panchromatic PL map acquired 4 µm above the NL. No localized emitters are observed at this depth over the same wavelength range: only defocused contributions from emitters located in the NL are detected.
(C) Panchromatic PL map acquired at the depth of the LT layer used to intentionally introduce emitters. Localized emitters are clearly observed, despite the measurement being performed 4 µm above the NL.
(D-F) Schematic representations of the sample structures probed in the measurements shown in (A-C), respectively.}\label{Fig_PL_maps}
\end{figure}

Figure \ref{Fig_PL_maps} (A) shows a panchromatic µ-PL map acquired at the depth of the NL of the 10 µm-thick sample. In this map, a high density of localized emitters is clearly resolved. To investigate the presence of emitters within bulk GaN, we performed similar measurements at depths away from the NL. Figure \ref{Fig_PL_maps} (B) presents a representative panchromatic map acquired 4 µm above the NL. Both PL maps were normalized to the highest detected intensity across the two measurements, which in this case belongs to the map acquired at the NL depth. No localized emitters are observed at this depth; only a weak, spatially extended background signal is detected, which originates from out-of-focus emitters located at the NL. This indicates the absence of any detectable emitters within the high-temperature (HT) GaN layers.

As a follow-up experiment, we performed identical PL mapping on a sample in which emitters were intentionally introduced 4 µm above the NL using a LT GaN interlayer (see Fig. 4 (A) in the main text). The corresponding panchromatic map (Fig. \ref{Fig_PL_maps} (C)) reveals localized emitters at this depth, demonstrating that the LT interlayer effectively reintroduces emitters with a density suitable for both isolation and characterization of individual QEs.

\section{SIMS measurements}

\begin{figure}[h!]
\centering
\includegraphics[width=0.55\textwidth]{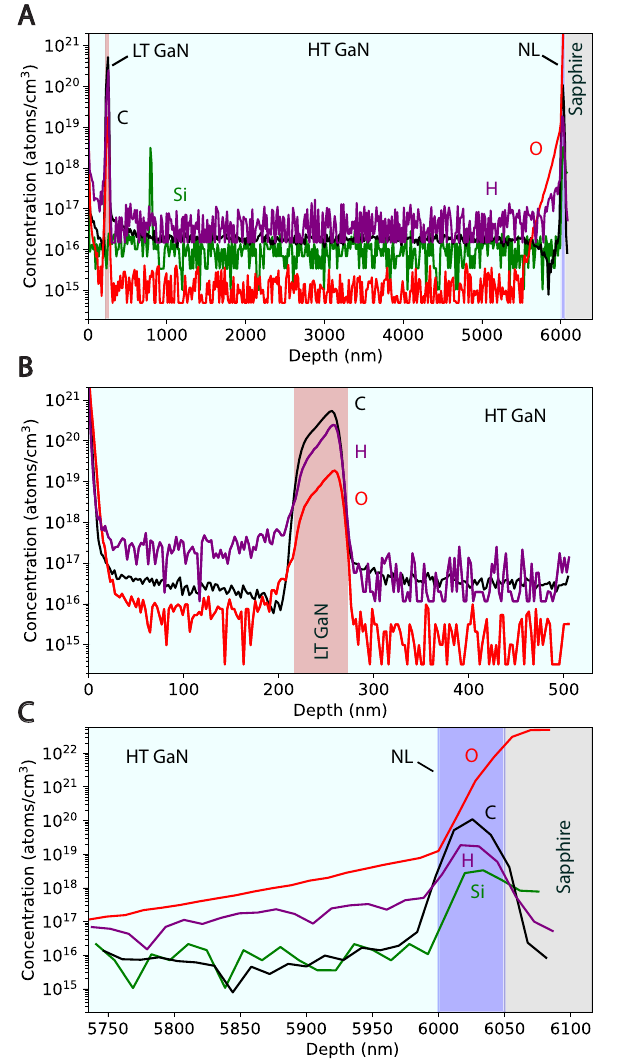}
\caption{(A) SIMS measurement for C, O, H and Si, all common impurities found in MOVPE-grown GaN. (B) High-resolution SIMS measurement around the LT layer containing intentionally introduced QEs. (C) Zoom-in on the NL region near the sapphire substrate, extracted from measurement (A).}\label{Fig_SIMS}
\end{figure}

The SIMS measurements were performed at EAG Laboratories and were used to determine the concentrations of common impurities in metalorganic vapor-phase epitaxy (MOVPE)-grown GaN. The measurements presented in Figs. \ref{Fig_SIMS} (A), (B) and (C) were performed on the sample presented in Fig. 5 of the main text, prior to any fabrication step. The sample consists of a 6-µm-thick GaN layer grown on sapphire, where a 30 nm LT GaN layer containing intentionally introduced QEs is positioned approximately 250 nm below the top surface. The sample was grown in two distinct growth steps. First, a standard 5.2-µm-thick GaN-on-sapphire template was grown. On top of this template, we carried out a regrowth where 550 nm of HT GaN were deposited, followed by the 30 nm LT GaN layer, and finally a 250 nm HT GaN cap. The result of the SIMS measurement across the whole epilayer is presented in Fig. \ref{Fig_SIMS} (A): within the HT GaN, the concentration of the main impurities found in MOVPE-grown GaN (C, O, H and Si) are mostly stable at low levels, and close to their detection limit for O, H and Si. The spike in Si concentration in the SIMS profile 800 nm below the surface is explained by the regrowth process, and we verified that the regrowth did not generate QEs on its own.

Higher levels in C, H and O concentrations appear when the GaN growth temperature is lowered, in the LT GaN layer and in the NL at the interface with the sapphire. A high-resolution SIMS measurement was recorded in the LT layer region for C, H and O (Fig. \ref{Fig_SIMS} (B)) revealing peak concentrations of $[\text{C}]_{\text{LT}} \simeq 5.5 \times 10^{20} \text{ cm}^{-3}$, $[\text{H}]_{\text{LT}} \simeq 2.5 \times 10^{20} \text{ cm}^{-3}$ and $[\text{O}]_{\text{LT}} \simeq 1.9 \times 10^{19} \text{ cm}^{-3}$. We quantify the depth spread of the impurity peaks to estimate the uncertainty in the vertical positioning of the introduced QEs. Although the peaks are asymmetric, their FWHM is approximately 20 nm. Using a more stringent criterion, we instead consider the depth range over which the impurity concentrations decrease by two orders of magnitude from their peak value. This range is approximately 58 nm, setting a lower bound on the positioning accuracy of the growth method. This value is in excellent agreement with the positioning accuracy determined from Fig. 5 of the main text. The impurity-containing region exceeds the nominal 30 nm thickness of the LT layer, likely owing to slight roughening during growth. The LT layer is also located at a slightly different depth than in Fig. 5 of the main text, as the measurements were performed at different positions on the sample.

A zoom-in around the NL region of Fig. \ref{Fig_SIMS} (A)  is presented in Fig. \ref{Fig_SIMS} (C), and reveals peak concentrations in the NL of $[\text{C}]_{\text{NL}} \simeq 1.1 \times 10^{20} \text{ cm}^{-3}$, $[\text{H}]_{\text{NL}} \simeq 1.8 \times 10^{19} \text{ cm}^{-3}$ and $[\text{O}]_{\text{NL}} \simeq 5-7 \times 10^{20} \text{ cm}^{-3}$. The proximity of the sapphire substrate and the depth of the measurement could influence the result of the SIMS, but we confirmed these values by performing a high-resolution SIMS measurement on the 800 nm-thick GaN on sapphire sample depicted in Fig. 3 (A) of the main text. An uncertainty however remains on the exact value of the O concentration in the NL due to the proximity of the sapphire substrate and the resulting steepness in O concentration in the NL. We chose to report the value measured at the depth where the concentrations of other impurities peak, which also coincides with the mean O concentration over the whole NL, but one has to keep in mind that there is an O concentration gradient in the NL. A summary of the impurity concentrations in the NL and the LT layer is presented in Table \ref{table_concentrations}, along with the detected emitter densities. The densities are estimated considering QEs emitting in the 520-780 nm spectral range.

\begin{table}[h!]
\centering
\caption{Comparison between impurity concentrations and QE densities.}\label{table_concentrations}%
\begin{tabular}{lll}
\\
\hline
 & LT GaN & GaN NL \\
\hline
C peak concentration & $5.5 \times 10^{20} \text{ cm}^{-3}$ & $1.1 \times 10^{20} \text{ cm}^{-3}$\\
H peak concentration &  $2.5 \times 10^{20} \text{ cm}^{-3}$ & $1.8 \times 10^{19} \text{ cm}^{-3}$ \\
O peak concentration & $1.9 \times 10^{19} \text{ cm}^{-3}$ & $5-7 \times 10^{20} \text{ cm}^{-3}$ \\
\hline
Emitter density & $2 \times 10^{7} \text{ cm}^{-2}$ & $3 \times 10^{8} \text{ cm}^{-2}$ \\
\hline
\end{tabular}
\end{table}

\section{Optical properties of introduced emitters}

\begin{figure}[h!]
\centering
\includegraphics[width=0.7\textwidth]{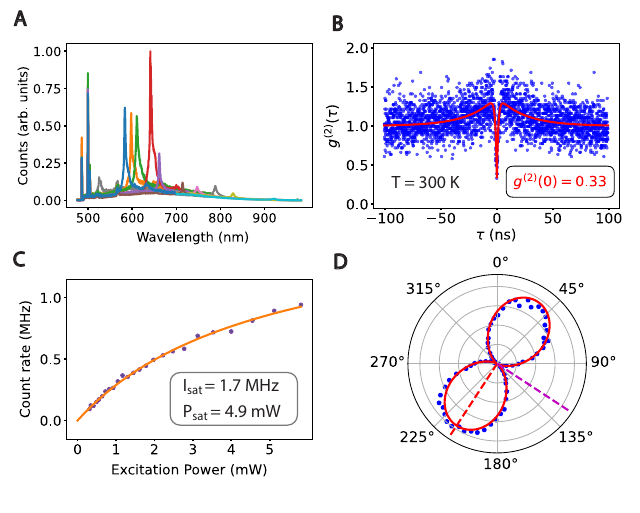}
\caption{(A) PL spectra of introduced QEs with emission wavelengths ranging from 530 to 920 nm. (B) Second-order correlation function measurement of an intentionally introduced emitter in the GaN layer. The excitation power was set to 100 µW in this measurement. (C) Power-dependent PL measurement of an introduced emitter. (D) Polarization-dependent measurement of the emitter characterized in (C), yielding a degree of linear polarization (DLP) of 97.1\%, and a main polarization axis forming a 213° angle with the direction perpendicular to the wurtzite $m$-plane.}\label{Fig_recreated}
\end{figure}

The QEs introduced at a chosen depth using a LT GaN layer exhibit optical properties very similar to those of emitters found in the NL. Specifically, defects with emission wavelengths spanning from 530 to 920 nm are observed (Fig. \ref{Fig_recreated} (A)), exhibiting antibunching at room temperature (Fig. \ref{Fig_recreated} (B)), high brightness (Fig. \ref{Fig_recreated} (C)), and linearly polarized emission (Fig. \ref{Fig_recreated} (D)). Additionally, the QEs exhibit narrow room-temperature linewidths, with FWHM values ranging from 3 to 9 nm (mean 5.4 nm), and high Debye-Waller factors between 0.69 and 0.98 (mean 0.81) obtained after background luminescence substraction. These observations strongly suggest that the intentionally introduced QEs originate from the same type(s) of defects as those naturally present in the NL.

We note that the signal-to-background ratio is lower for some intentionally introduced emitters than for emitters in the NL, which accounts for the higher \textit{g}$^{(2)}(0)$ value presented in Fig. \ref{Fig_recreated} (B) compared to that of a naturally occurring emitter in GaN (Fig. 1 (C) in the main text). The broadband background luminescence observed in all spectra, the so-called yellow luminescence band in GaN, is commonly attributed to C$_{\text{N}}$ defects, corresponding to carbon substitution on nitrogen sites \cite{reshchikov_origin_2023}. Further fine-tuning of the growth process is thus expected to improve the signal-to-background ratio.

\FloatBarrier

\section{Emitters within ridges}

\begin{figure}[h!]
\centering
\includegraphics[width=1.\textwidth]{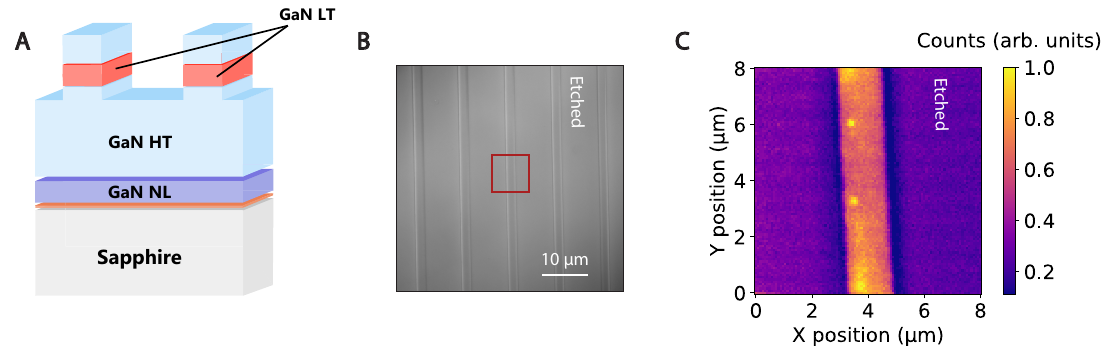}
\caption{(A) Schematic of the sample structure with fabricated ridges. (B) Wide-field image of the etched ridges. The red box displays the spatial extent of the PL map shown in (C). (C) PL map of a ridge. The map is plotted at the emission wavelengths of two distinct emitters and superimposed with the map of the Raman $E_2$ (high) mode to highlight the ridge.}\label{Fig_ridges}
\end{figure}

Figure \ref{Fig_ridges} presents results obtained from the sample shown in Fig. 5 of the main text. We patterned 2 µm-wide ridges containing the emitter-forming layer with a single photolithography step, while etching it away outside of the ridges (Fig. \ref{Fig_ridges} (A)). Emitters were thus expected to be detected exclusively within the ridges. Figure \ref{Fig_ridges} (B) shows the sample after nanofabrication, imaged with our home-built setup, while Fig. \ref{Fig_ridges} (C) presents the µ-PL measurements over a representative ridge. The focusing depth was set just below the etched surface to probe the GaN both inside the ridge and outside the ridge. In this configuration, the introduced emitters were well within the depth of focus of the objective, whereas QEs at the GaN/sapphire interface were completely out of focus and thus not detected. To correlate the detected emitter positions with the ridge geometry, we overlaid the PL map plotted at the emission wavelengths of two bright emitters and at the Raman $E_2$ (high) wavelength. When scanning over the ridge, more GaN material was probed, resulting in a higher Raman signal and causing the ridge to appear brighter in the map. All detected emitters, as expected, were located inside the ridge. In this PL map, even though a larger area was scanned outside the ridge, 7 emitters were detected within the ridge and none outside. Additional PL maps confirmed the absence of emitters outside the ridges. From the mapped area and the lack of detected emitters, we infered an upper bound on the emitter density that is at least three orders of magnitude lower than inside the ridges. These results reinforce the conclusions drawn from Fig. 5 of the main text.

\bibliography{depth_control_si}